\documentclass[aps,showpacs]{revtex4}
\usepackage{amsmath}
\usepackage{amssymb}
\usepackage{graphicx}% Include figure files
\usepackage{dcolumn}% Align table columns on decimal point
\usepackage{bm}% bold math

\newcommand{\la}{\lambda}
\newcommand{\ga}{\gamma}

\newcommand{\om}{\omega}

\newcommand{\br}{{\bf r}}
\newcommand{\prt}{\partial}
\newcommand{\bu}{{\bf u}}
\newcommand{\bk}{{\bf{k}}}

\newcommand{\txi}{{\tilde{\xi}}}
\newcommand{\teta}{{\tilde{\eta}}}
\newcommand{\tro}{{\tilde{\rho}}}
\newcommand{\tz}{{\tilde{z}}}
\newcommand{\tx}{{\tilde{x}}}
\newcommand{\ty}{{\tilde{y}}}
\newcommand{\psd}{\psi^{\prime\prime}}

\begin{document}

\title{
Nonlinear diffraction of light beams propagating in photorefractive
media with embedded reflecting wire}

\author{E.G. Khamis$^{1}$}
\email{egkhamis@if.usp.br}
\author{A. Gammal$^{1}$}
\email{gammal@if.usp.br}
\author{G.A. El$^{2}$}
\email{G.El@lboro.ac.uk}
\author{Yu.G. Gladush$^{3}$}
\email{gladush@isan.troitsk.ru}
\author{A.M. Kamchatnov$^{3}$}
\email{kamch@isan.troitsk.ru}

\affiliation{
$^1$ Instituto de F\'{\i}sica, Universidade de S\~{a}o Paulo,
05315-970, C.P.66318 S\~{a}o Paulo, Brazil\\
$^2$ Department of Mathematical Sciences, Loughborough
University, Loughborough LE11 3TU, UK \\
$^3$Institute of Spectroscopy, Russian Academy of Sciences, Troitsk,
Moscow Region, 142190, Russia\\ }

\date{\today}

\begin{abstract}
The theory of nonlinear diffraction of intensive light beams
propagating through photorefractive media is developed. Diffraction
occurs on a reflecting wire embedded in the nonlinear medium at
relatively small angle with respect to the direction of the beam
propagation. It is shown that this process is analogous to the
generation of waves by a flow of a superfluid past an obstacle. The
``equation of state'' of such a superfluid is determined by the
nonlinear properties of the medium. On the basis of this
hydrodynamic analogy, the notion of the ``Mach number'' is
introduced where the transverse component of the wave vector
plays the role of the fluid velocity. It is found that the Mach cone
separates two regions of the diffraction pattern: inside the Mach
cone oblique dark solitons are generated and outside the Mach cone
the region of ``ship waves'' is situated. Analytical theory of
``ship waves'' is developed and two-dimensional dark soliton
solutions of the equation describing the beam propagation are found.
Stability of dark solitons with respect to their decay into vortices
is studied and it is shown that they are stable for large enough
values of the Mach number.
\end{abstract}


\maketitle

\section{Introduction}

An analogy between propagation of light beams in nonlinear media and
superfluid flow is well known and quite suggestive. Formally, it is
based on a mathematical similarity of the equations for
electromagnetic field evolution of light beams in paraxial
approximation and Gross-Pitaevskii equations for superfluid motion
of Bose-Einstein condensates of dilute gases. Accordingly, such
nonlinear structures as bright or dark solitons and vortices have
been thoroughly studied both in optics and superfluid dynamics (see,
e.g., \cite{ka03,ps03}). These structures arise as a results of
interplay of nonlinear and dispersive properties of the medium under
consideration. One more example of such a structure is provided by
so-called dispersive shocks which replace a notion of usual
dissipative shocks in compressive fluid dynamics in case when
dissipation can be neglected compared with dispersive effects. As a
result, a thin layer with strong dissipation within unfolds into a
region with fast oscillations, which can be represented as a
modulated nonlinear periodic wave (a ``soliton lattice''). The
notion of dispersive shocks arose first in water wave physics (the
theory of undular bores on rivers) \cite{bl54} and plasma physics
(collisionless shock waves) \cite{sagdeev}, then generality of
this phenomenon was realized and (based on the Whitham theory
\cite{whitham1} of modulations of nonlinear waves) mathematical
methods for their description were developed \cite{GP1}-\cite{el05}.

Realization of Bose-Einstein condensate of dilute cold gases
\cite{bec1,bec2,bec3} and study of its dynamics has naturally led to
the theoretical and experimental studies of dispersive shocks in
this new medium \cite{damski04}--\cite{engels07}. Corresponding
optical counterpart of dispersive shocks suggested by the above
mentioned analogy between beam optics and superfluid dynamics was
realized experimentally in
\cite{fleischer,trillo,fleischer3,fleischer2} and the theory of such
optical dispersive shocks was developed in \cite{el07}.

In dissipative fluid dynamics with negligible dispersion shocks can
also be generated by a supersonic flow of the fluid past an
obstacle. Such shocks have the form of a sharp stationary jump
of the fluid parameters across certain lines inclined with respect
to the flow direction. For shocks of small intensity these lines lie
along the so-called ``Mach cones'' (see, e.g. \cite{LL6}). In
dispersive fluid dynamics these oblique shocks unfold into
``fans'' of spatial solitons spreading downstream from the obstacle
\cite{karpman}. The theory of such oblique dispersive shocks
was developed in \cite{GKKE95} for the case of weakly
dispersive media when the
 flow past a slender body is asymptotically described by the
Korteweg-de Vries equation along the Mach lines.

Dynamics of a Bose-Einstein condensate is described by the
Gross-Pitaevskii equation and the theory was extended to this case
in \cite{ek06}. If the obstacle is small enough, then the shock
consists of a single oblique dark soliton. The theory of
oblique dark solitons was developed in \cite{egk06,kp07}. It is
important to note that such oblique solitons are located inside the
Mach cone with the Mach number defined as a ratio of the flow
velocity to the sound speed calculated at infinite wavelength. The
so-called ``ship waves'' arising as stationary dispersive wave
packets of Bogoliubov excitations are located outside the Mach cone.
Apparently, they were observed in the experiment \cite{caruso} and
their theory was developed in \cite{gegk07,gk07}. The analogy
between beam optics and superfluid dynamics suggests that similar
effects would exist in the optical context where they
take the form of diffraction wave patterns in light
beams propagating through a nonlinear medium. Although such
structures were observed in some experiments (see, e.g.,
\cite{sl92}), they have not been studied systematically yet. In this
paper, we shall consider a typical simple situation of nonlinear
diffraction of light which can be considered as an optical
counterpart of generation of spatial dispersive shocks and
``ship waves'' in the flow of Bose-Einstein condensate past an
obstacle. To be definite, we consider a light beam propagating
through a bulk self-defocusing nonlinear refractive medium with a
thin wire (a ``needle'') inserted in it; see Fig.~1. Direction of
the light beam is tilted with respect to the wire that is there
exists a ``flow'' of light ``past an obstacle''. As a result, at the
output plane of the medium a diffraction pattern is formed
consisting of oblique dark solitons and ``ship waves''. We shall
give here analytical and numerical treatment of this phenomenon and
obtain main characteristics of the diffraction pattern.

\begin{figure}[bt]
\begin{center}
\includegraphics[width=8cm,height=8cm,clip]{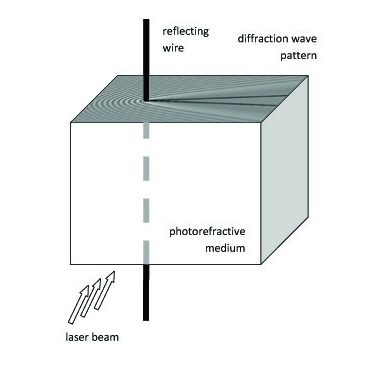}
\caption{A sketch of formation of nonlinear diffraction pattern in
propagation of a light beam through photorefractive medium with
embedded reflecting wire.}
\end{center}\label{fig1}
\end{figure}

\section{Main equations and general form of the diffraction pattern}

Propagation of stationary beams is described by the
equation
\begin{equation}\label{1-1}
    i\frac{\prt\psi}{\prt z}+\frac1{2k_0}\Delta_\bot\psi
    +\frac{k_0}{n_0}
    \delta n\left(|\psi|^2\right)\psi +V(\br)\psi=0,
\end{equation}
where $\psi$ is envelope field strength of electromagnetic wave with
wave number $k_0=2\pi n_0/\la$, $z$ is the coordinate along the
beam, $x,y$ are transverse coordinates, $\br =(x,y)$,
$\Delta_\bot=\prt^2/\prt^2x+\prt^2/\prt^2y$ is transverse Laplacian,
$n_0$ is a linear refractive index, $V(\br)$ represents a
``potential'' of an obstacle (e.g. a reflecting wire) at which
diffraction occurs, and in a photo-refractive medium we have
\begin{equation}\label{1-2}
    \delta n=-\frac12 n_0^3 r_{33}E_p\frac{\rho}{\rho+\rho_d},
\end{equation}
where $E_p$ is applied electric field, $r_{33}$ electro-optical
index, $\rho=|\psi|^2$, and $\rho_d$ is the saturation parameter.

For mathematical convenience, we introduce non-dimensional variables
\begin{equation}\label{1-4}
    \tilde{z}=\frac12 kn_0^2r_{33}E_p\left(\frac{\rho_c}{\rho_d}\right) z,\quad
    \tilde{x}=kn_0\sqrt{\frac12r_{33}E_p\left(\frac{\rho_c}{\rho_d}\right)}x,\quad
    \tilde{y}=kn_0\sqrt{\frac12r_{33}E_p\left(\frac{\rho_c}{\rho_d}\right)}y,\quad
    \tilde{\psi}=\sqrt{\rho_c}\psi,
\end{equation}
where $\rho_c$ is a characteristic value of optical intensity (its
concrete definition depends on the problem under consideration; for
instance, it can be the background intensity), so that
Eq.~(\ref{1-1}) takes the form of generalized nonlinear
Schr\"odinger (GNLS) equation
\begin{equation}\label{1-5}
    i\frac{\prt\psi}{\prt z}+\frac1{2}\Delta_\bot\psi-
    \frac{|\psi|^2}{1+\gamma|\psi|^2}\psi +V(\br)\psi=0,
\end{equation}
where $\gamma=\rho_c/\rho_d$, $V(\br)$ is represented in
non-dimensional units, and tildes are omitted for convenience of the
notation. In fact, our approach can be applied to other forms of the
nonlinear term provided it corresponds to self-defocusing light
beams. Therefore we shall also use  the general form of the equation
\begin{equation}\label{1-5a}
    i\frac{\prt\psi}{\prt z}+\frac1{2}\Delta_\bot\psi-
    f(|\psi|^2)\psi +V(\br)\psi=0,
\end{equation}
 where $f(|\psi|^2)>0$, In particular, for photorefractive medium,
\begin{equation}\label{1-5b}
f(\rho)=\rho/(1+\gamma\rho).
\end{equation}

If saturation effect is negligibly small ($\gamma|\psi|^2\ll1$),
then Eq.~(\ref{1-5}) reduces to the standard cubic nonlinear
Schr\"odinger (NLS) equation
\begin{equation}\label{1-6}
    i\frac{\prt\psi}{\prt z}+\frac1{2}\Delta_\bot\psi-
    {|\psi|^2}\psi+V(\br)\psi=0.
\end{equation}

If the phase of $\psi$ is a single-valued function, then it is
convenient to represent the above NLS equations in a fluid dynamics
type form by means of the substitution
\begin{equation}\label{1-7}
    \psi(\br, z)=\sqrt{\rho}\,\exp\left(i \int^{\br}
    \bu(\br´,z) \cdot d\br´\right) ,
\end{equation}
so that they are transformed into
\begin{equation}\label{1-8}
\begin{split}
\rho_z+\nabla_\bot(\rho \bu)=0, \\
\bu_z+\left(\bu\nabla_\bot\right)\bu +\nabla_\bot f(\rho)-\nabla V(\br)-\nabla_\bot
\left[\frac{\Delta_\bot\rho}{4\rho}-\frac{(\nabla_\bot\rho)^2}{8\rho^2}
\right] =0.
\end{split}
\end{equation}
In the hydrodynamic interpretation the light intensity $\rho$ has a
meaning of a density of a ``fluid'' and Eq.~(\ref{1-5b}) can be
viewed as an ``equation of state'' for such a fluid. The function
$\bu(\br,z)$ is a local value of the wave vector component
transverse to the direction of the light beam; in hydrodynamic
representation it has a meaning of the ``flow velocity''. The
variable $z$ plays the role of time so it is natural to describe the
deformations of the light beam in evolutionary terms.  We note that
substitution (\ref{1-7}) rules out vorticity so that system
(\ref{1-8}) actually represent a restriction of the
multi-dimensional GNLS equation (\ref{1-5a}) to potential ``flows''.

We shall consider propagation of a tilted light beam with uniform
input intensity, that is at $z=0$ it has the initial form
\begin{equation}\label{2-2}
    \psi(\br,0)=\exp(iUx),
\end{equation}
that is we suppose that the background intensity is equal to unity;
$U$ represents the $x$-component of the  wave vector due to tilting
of the light beam. The problem is to describe the wave pattern at
the output value of $z$.

To clarify a general picture of the
diffraction pattern, we have solved Eq.~(\ref{1-5}) numerically
for the initial wave function $\psi$ given by Eq.~(\ref{2-2})
with $U=2$ and the boundary condition of vanishing $\psi$ at
the surface $r=1$ of the obstacle located at $x=0,\,y=0$.
\begin{figure}[bt]
\includegraphics[width=8cm,height=6.5cm,clip]{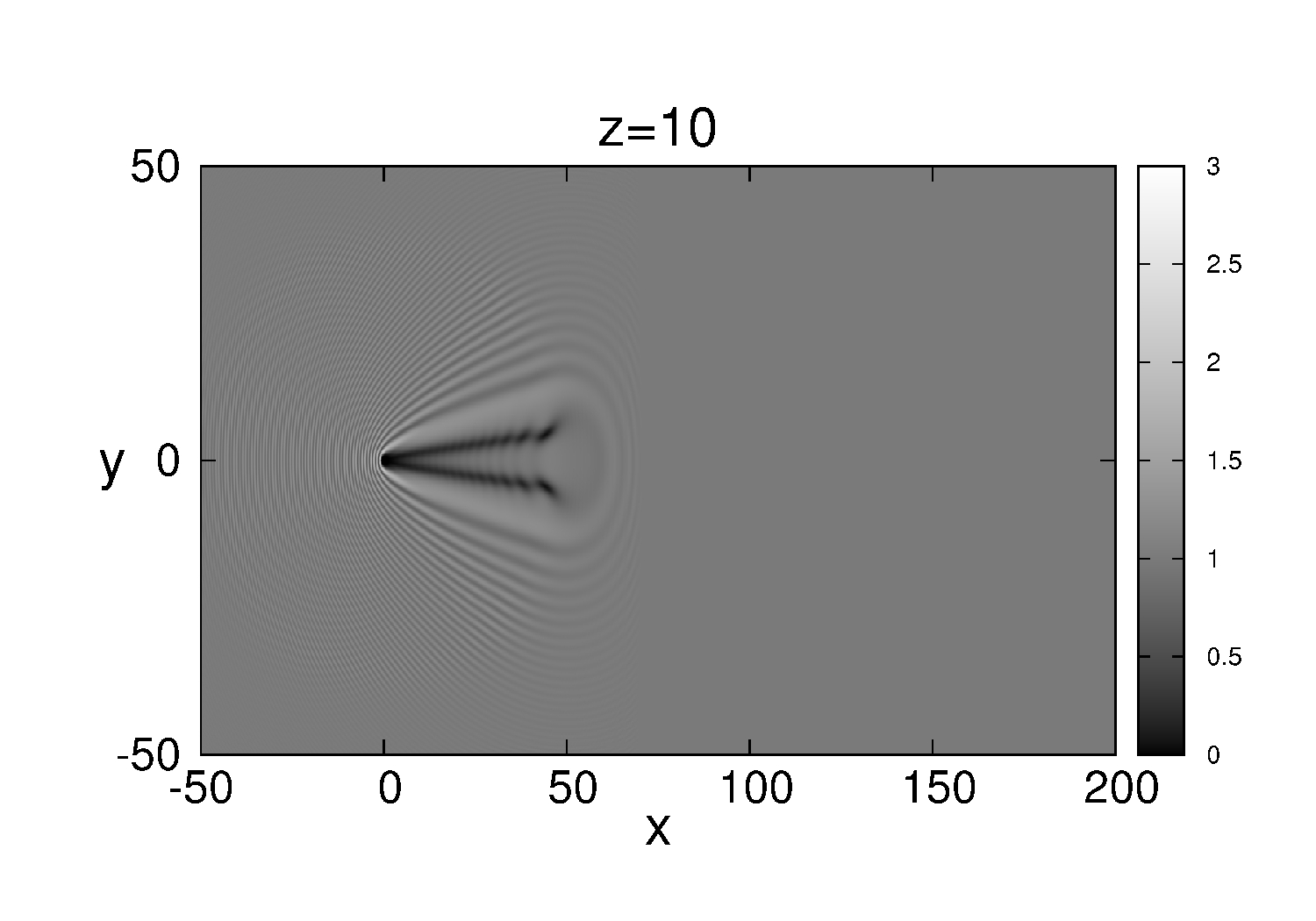}
\includegraphics[width=8cm,height=6.5cm,clip]{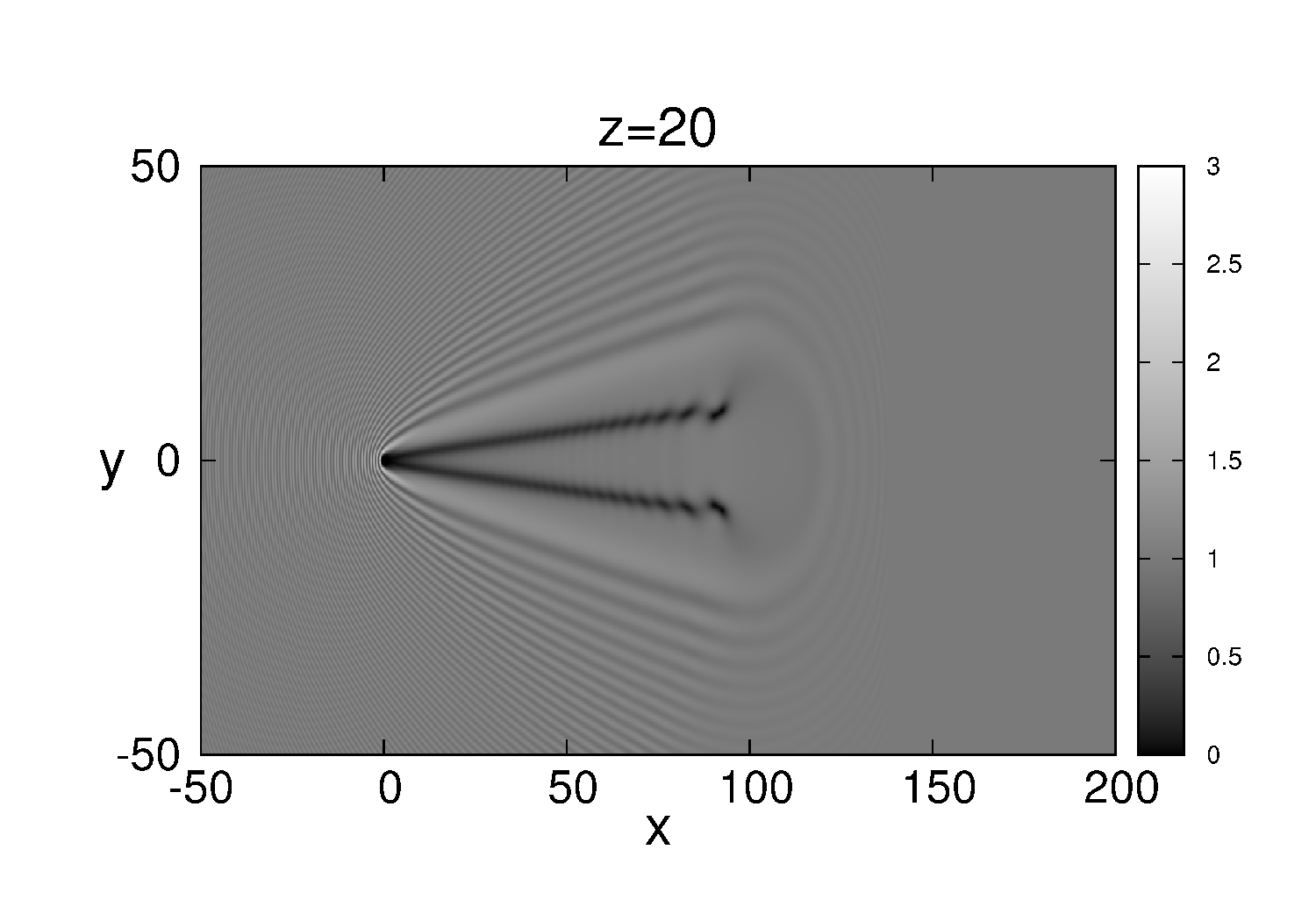}
\includegraphics[width=8cm,height=6.5cm,clip]{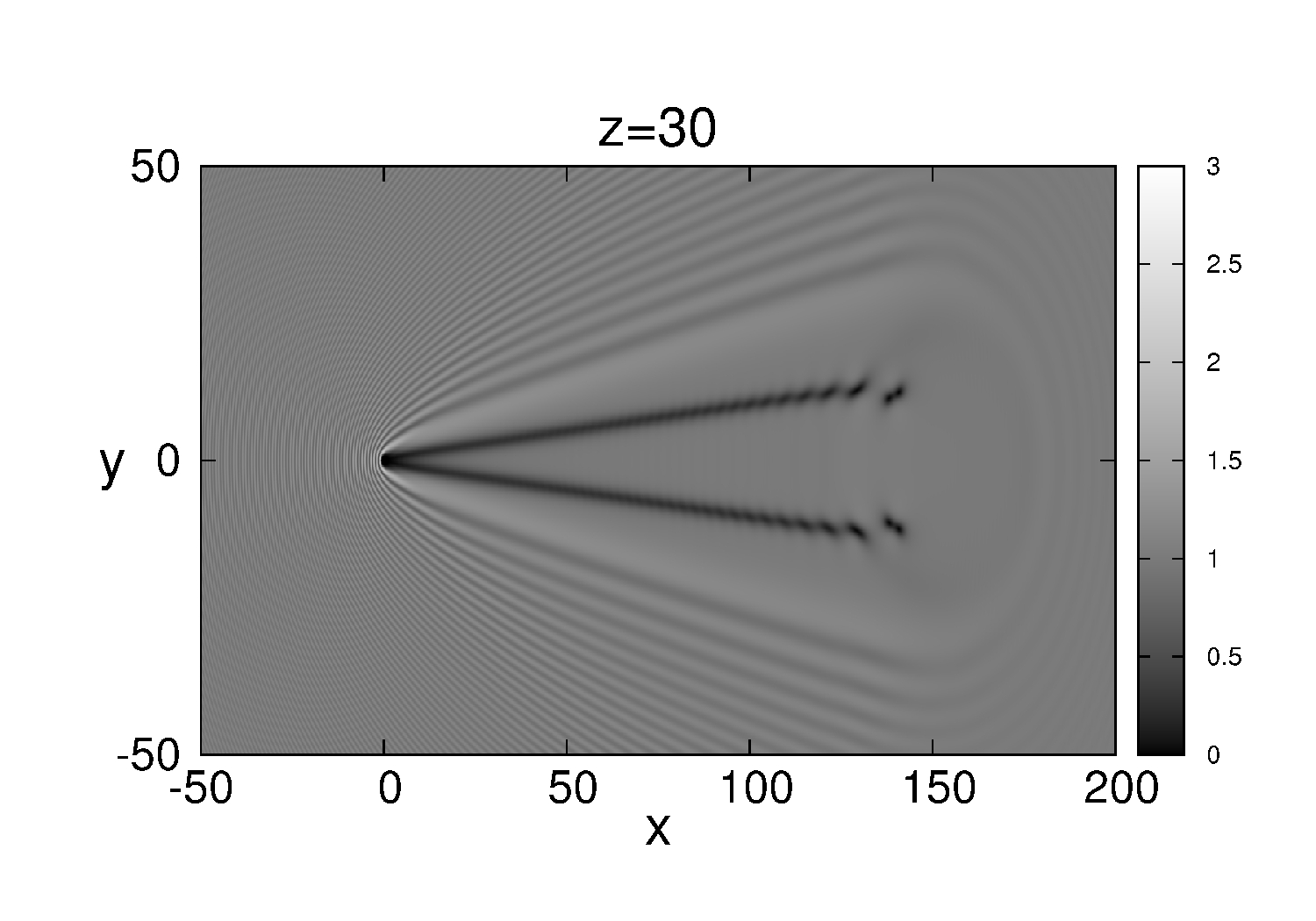}
\caption{Evolution of the diffraction pattern at the output plane
as a function of the length $z$ of the photorefractive medium.
The patterns are obtained by numerical solution of Eq.~(\ref{1-5})
with $V(\br)$ corresponding to an ideally reflecting wire with
unit radius for $\ga=0.2$, $U=2$, and (a) $z=20$, (b) $z=40,$
(c) $z=60$.
} \label{fig2}
\end{figure}
As we see, the diffraction pattern consists of two different parts
separated by the Mach (or Cherenkov) cone which is defined as lines
drawn  at angle $\theta$ with respect to the direction of the flow ($x$ axis)
with
\begin{equation}\label{mach}
    \sin\theta=\frac1M,\qquad M=\frac{U}{c_s}
\end{equation}
where the sound velocity corresponds to the dispersionless limit of
Eqs.~(\ref{1-8}) that is ($\nabla p/\rho\equiv \nabla f(\rho)$)
\begin{equation}\label{sound1}
    c_s=\left.\sqrt{\frac{dp}{d\rho}}\right|_{\rho_0}=\sqrt{f'(\rho_0)\rho_0}
\end{equation}
which in the photorefractive case with $\rho_0=1$ yields
\begin{equation}\label{sound2}
    c_s=\frac1{1+\ga}\quad \mathrm{and}\quad M=U(1+\ga).
\end{equation}
Outside the Mach cone, there is a stationary wave pattern created by
interference of linear (far enough from the obstacle) waves. Inside
the Mach cone there are two oblique dark solitons situated
symmetrically with respect to the direction of the ``flow''. These
oblique solitons decay at the end points into vortices but closer to
the obstacle they are described by a potential flow with jump of
phase across them as it is demonstrated in Fig.~3.
\begin{figure}[bt]
\includegraphics[width=8cm,height=7cm,clip]{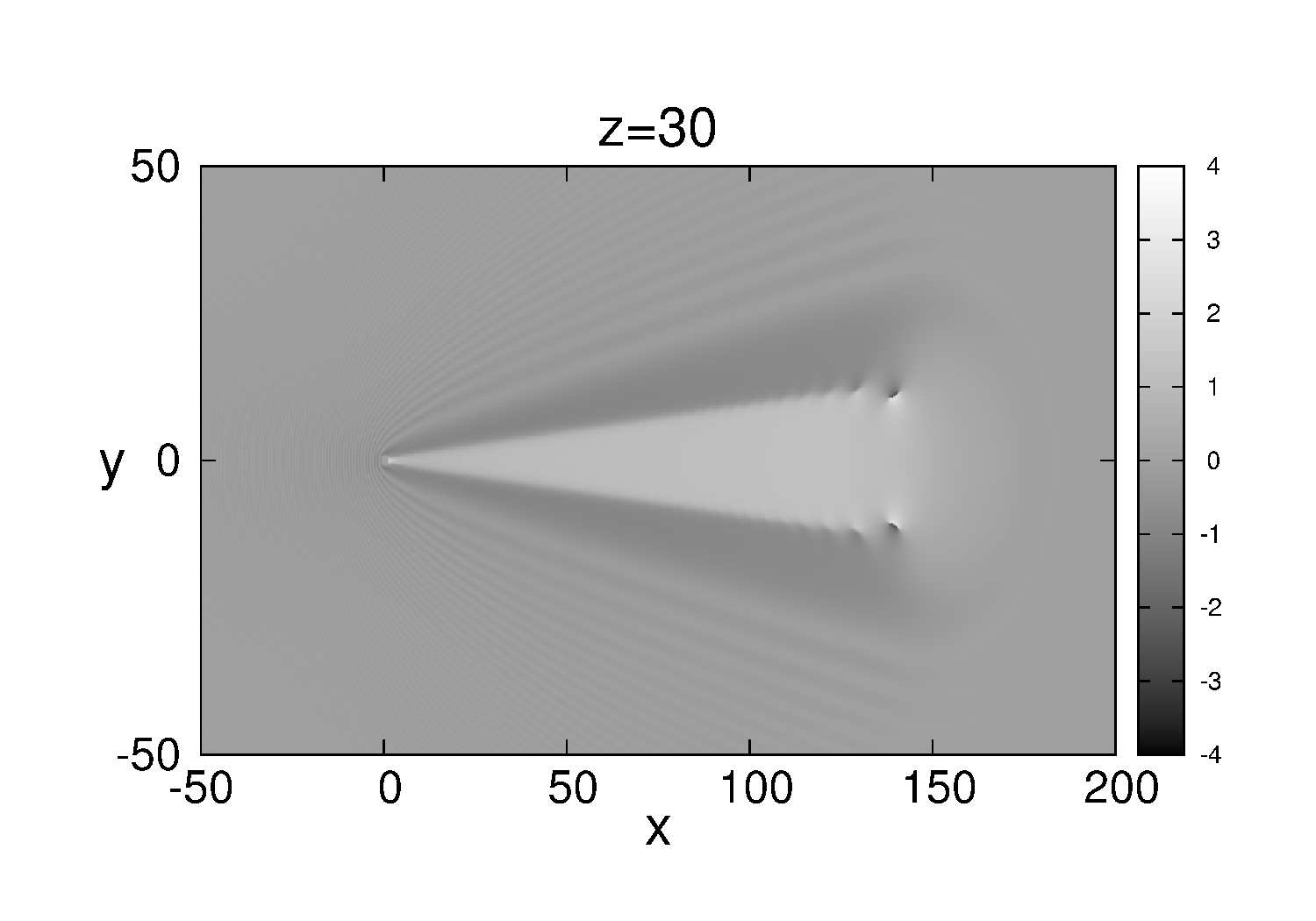}
\caption{Distribution of the phase in the diffraction pattern at the output plane
of the photorefractive medium.
The pattern corresponds to $\ga=0.2$, $U=2$, and $z=60$.
} \label{fig3}
\end{figure}

Our task now is to develop analytical theory for these two regions of
the diffraction pattern and to compare it with numerical simulations.
We shall start with the ``ship waves'' pattern located outside the Mach cone.

\section{Diffraction pattern outside the Mach cone}

If the size of the obstacle is much less than the wavelength of the pattern,
then we can consider it as a point-like one and take the obstacle potential
in the form
\begin{equation}\label{pot}
    V(\mathbf{r})=V_0\delta(\mathbf{r}).
\end{equation}
Far enough from the obstacle, the amplitude of the wave pattern is
small compared with the background intensity of the light beam.
Hence, the wave pattern can be calculated by means of perturbation
theory \cite{ap-04}.

If we neglect the influence of the obstacle, then the
$\psi$-function of a uniform light beam with the intensity $\rho_0$
depends on $z$ in the reference frame with $U=0$ as
$\psi\propto\exp(-if(\rho_0)z)$. We exclude this dependence by
introducing the substitution $\psi=\Psi\cdot\exp(-if(\rho_0)z)$ so
that $\Psi$ satisfies the equation
\begin{equation}\label{GP-2}
    i\Psi_z+\tfrac12\Delta\psi +\left[f(\rho_0)-f(|\Psi|^2)\right]\Psi=0.
\end{equation}
In the same reference frame the obstacle moves with the velocity $\mathbf{-U}$
and generates diffraction waves which in the linear approximation
are described by a small correction $\delta\Psi$
to the unperturbed wave function: $\Psi\approx \sqrt{\rho_0}+\delta\Psi$.
Hence $\delta\Psi$ satisfies the equation
\begin{equation}\label{pert-1}
    i\delta\Psi_z+\tfrac12\Delta\delta\Psi-c_s^2(\delta\Psi+\delta\Psi^*)-
    V_0\sqrt{\rho_0}\delta(\mathbf{r}+\mathbf{U}z)=0
\end{equation}
where we have added the potential of the obstacle due to which
linear waves are generated. In the stationary case, which we are
interested in, the wave pattern moves with the obstacle, that is in
the reference frame attached to the reflecting wire we have
$\Psi=\Psi(\mathbf{r}+\mathbf{U}z)$ and
$$
\frac{\prt}{\prt z}\delta\Psi(\mathbf{r}+\mathbf{U}z)=
(\mathbf{U}\nabla)\delta\Psi(\mathbf{r}+\mathbf{U}z).
$$
Introducing $\mathbf{r'}=\mathbf{r}+\mathbf{U}z$ and omitting
primes, we arrive at the equation
\begin{equation}\label{pert-2}
    i(\mathbf{U}\nabla)\delta\Psi+\tfrac12\Delta\delta\Psi-c_s^2(\delta\Psi+\delta\Psi^*)-
    V_0\sqrt{\rho_0}\delta(\mathbf{r})=0\, ,
\end{equation}
describing stationary diffraction pattern generated by the beam.

Equation (\ref{pert-2}) can be solved by the Fourier method. We introduce the Fourier
transform of the wave function:
\begin{equation}\label{fourier}
    \delta\Psi=\int\delta\Psi_{\mathbf{k}}e^{i\mathbf{k}\mathbf{r}}\frac{d^2k}{(2\pi)^2},\quad
    \delta\Psi^*=\int\delta\Psi_{\mathbf{k}}^*e^{-i\mathbf{k}\mathbf{r}}\frac{d^2k}{(2\pi)^2}
\end{equation}
and obtain
\begin{equation}\label{eq1}
    -(\mathbf{k}\mathbf{U}+k^2/2+c_s^2)\delta\Psi_{\mathbf{k}}-c_s^2\delta\Psi_{-\mathbf{k}}^*=V_0\sqrt{\rho_0}.
\end{equation}
Another equation is obtained by means of substitution $\mathbf{k}\to-\mathbf{k}$ and complex
conjugation:
\begin{equation}\label{eq2}
    -c_s^2\delta\Psi_{\mathbf{k}}+(\mathbf{k}\mathbf{U}-k^2/2-c_s^2)\delta\Psi_{-\mathbf{k}}^*=V_0\sqrt{\rho_0}.
\end{equation}
Solution of Eqs.~(\ref{eq1},\ref{eq2}) reads
\begin{equation}\label{sol1}
    \delta\Psi_{\mathbf{k}}=V_0\sqrt{\rho_0}\frac{k^2/2-\mathbf{k}\mathbf{U}}{(\mathbf{k}\mathbf{U})^2-k^2(c_s^2+k^2/4)}.
\end{equation}
Since
$$
\delta
\rho=\sqrt{\rho_0}(\delta\Psi+\delta\Psi^*)=\int(\delta\Psi_{\mathbf{k}}+\delta\Psi_{-\mathbf{k}}^*)
e^{i\mathbf{k}\mathbf{r}} \frac{d^2k}{(2\pi)^2}\,
$$
we arrive at the following expression for the intensity perturbation in the output diffraction wave pattern
created by propagation of light past a reflecting wire:
\begin{equation}\label{2-9}
    \delta \rho=V_0\rho_0\int\frac{k^2e^{i\mathbf{k}\mathbf{r}}}{(\mathbf{k}\mathbf{U})^2-k^2(c_s^2+k^2/4)+i0}
    \frac{d^2k}{(2\pi)^2}\, ,
\end{equation}
where we have introduced an infinitesimal positive imaginary term $+i0$ corresponding to
the radiation condition for outgoing waves.

\begin{figure}[bt]
\begin{center}
\includegraphics[width=6cm,height=6cm,clip]{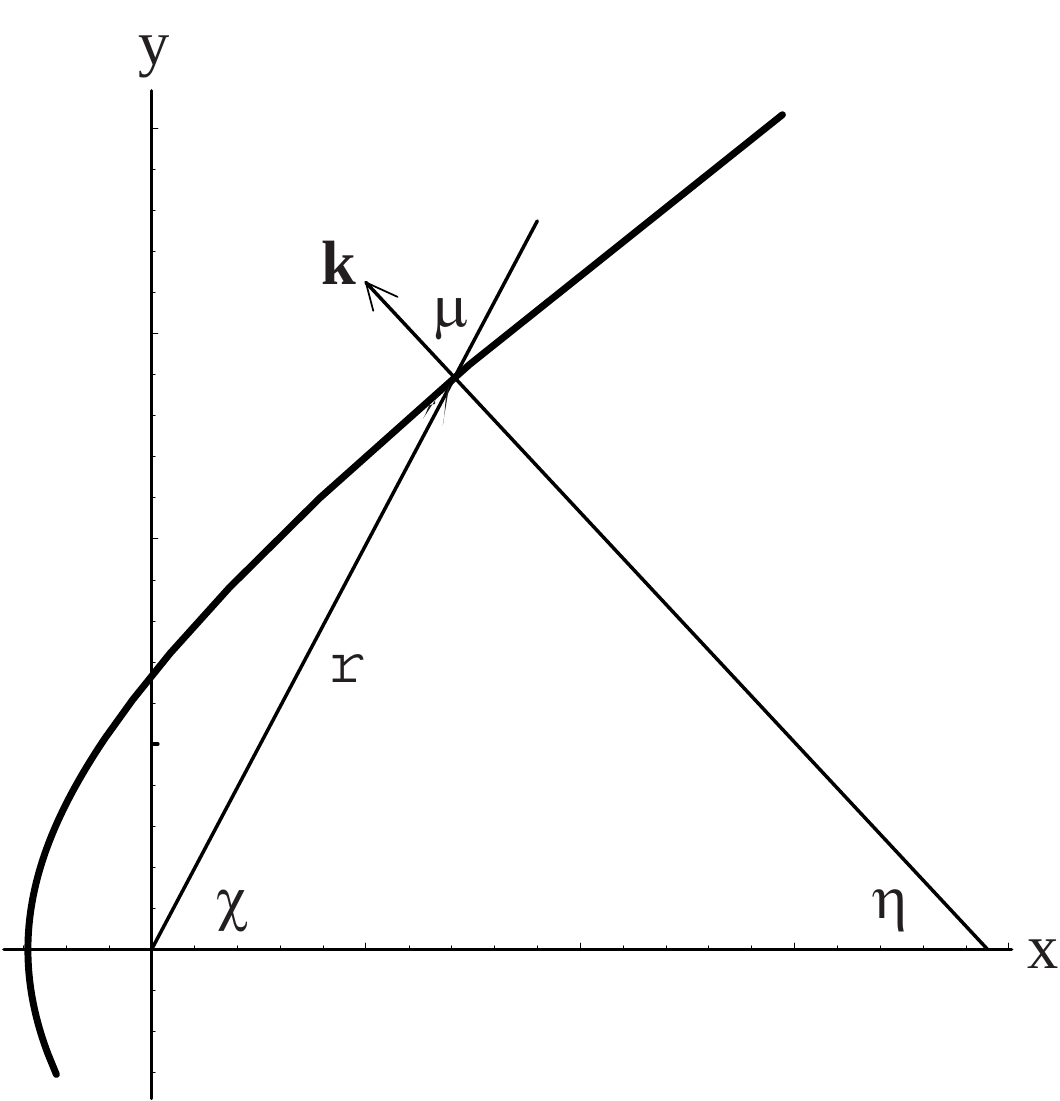}
\caption{Coordinates defining a radius vector $\br$ and a wave
vector $\bk$, normal to the wave front shown schematically by a
curved line. }
\end{center}\label{fig4}
\end{figure}

Now we introduce polar coordinates (see Fig.~4) defining the
components of the vectors $\br$ and $\bk$ as
\begin{equation}\label{2-11}
\begin{split}
x=r\cos{\chi}, \quad y=r\sin{\chi}; \\ \qquad k_x=-k\cos{\eta},\quad
k_y=k\sin{\eta}.
\end{split}
\end{equation}
Simple transformation casts Eq.~(\ref{2-9}) to the form
\begin{equation} \label{2-12}
\delta \rho=\frac{V_0\rho_0}{\pi^2}\int_{-\pi}^{\pi}\int_{0}^{\infty}
\frac{ke^{-ikr\cos(\chi+\eta)}d k d \eta}{k^2-k_0^2-i0},
\end{equation}
where
\begin{equation}\label{2-13}
k_0=2c_s\sqrt{M^2\cos^2{\eta}-1}=c_s\widetilde{k}(\eta).
\end{equation}
We can represent the integral (\ref{2-12}) as a sum
$$
\int_{-\pi/2}^{3\pi/2}
d\eta=\int_{-\pi/2}^{\pi/2} d\eta+\int_{\pi/2}^{3\pi/2} d\eta
$$
and,
noticing that the second term after substitution $\eta'=\eta-\pi$
becomes equal to a complex conjugate of the first one, we rewrite it as
\begin{equation} \label{2-14}
\delta\rho=\frac{V_0\rho_0}{\pi^2}\mathrm{Re}\int_{-\pi/2}^{\pi/2}\int_{0}^{\infty}
\frac{ke^{-ikr\cos(\chi+\eta)}d k d \eta}{k^2-k_0^2-i0}.
\end{equation}
To perform integration over $k$, we notice that the integrand function
has a pole in the first quadrant,
\begin{equation}\label{2-15}
k=\sqrt{k_0^2+i0}= k_0+{i0},
\end{equation}
which gives the main contribution into the integral for
$\cos(\chi+\eta)<0$. Indeed, taking a closed contour along the
positive real axis of $k$ with added quarter of the circle, which
gives no contribution into the integral, and a path along positive
imaginary axis which contribution
\begin{equation}\label{2-16}
\int_{0}^{\infty}\frac{k e^{-kr\cos(\chi+\eta)}d
k}{k^2+k_0^2}\propto\frac1{r^2},
\end{equation}
is decreasing with $r$ much faster than the contribution of the pole (which is proportional to
$r^{-1/2}$; see below), we obtain
\begin{equation} \label{2-17}
\delta \rho=-\frac{2V_0\rho_0}{\pi}\,\mathrm{Im}\int_{-\pi/2}^{\pi/2}e^{-ikr\cos(\chi+\eta)}d\eta,
\end{equation}
where $k$ is determined by the equation (\ref{2-13}) (index ``0''
is omitted here).

If the phase $\bk\br=r\varphi$, where
\begin{equation}\label{2-18}
    \varphi(\eta)=k(\eta)\cos(\chi+\eta),
\end{equation}
is large enough, the integral (\ref{2-17}) can be evaluated by the
standard method of stationary phase. This condition is fulfilled far
enough from the obstacle $r\rightarrow\infty$ provided
$|k(\eta)\cos(\chi+\eta)|\gg 1/r$. The equation which determines the
point of the stationary phase $\prt \varphi/\prt\eta=0$ gives
relationships for the angles (see Fig.~4)
\begin{equation}\label{2-19}
\begin{split}
    \tan\mu=\frac{2U^2}{k^2}\sin2\eta=\frac{2M^2}{\widetilde{k}^2}\sin2\eta, \quad
    \tan\chi=\frac{(c_s^2+k^2/2)\tan\eta}{U^2-(c_s^2+k^2/2)}=\frac{(1+\widetilde{k}^2/2)\tan\eta}
    {M^2-(1+\widetilde{k}^2/2)}.
\end{split}
\end{equation}
Taking into account equation (\ref{2-13}), we find
\begin{equation} \label{7-6}
\cos{\mu}=\frac{\widetilde{k}^2}{2[(M^2-2)\widetilde{k}^2+4(M^2-1)]^{1/2}}.
\end{equation}
With account of (\ref{2-19}), we get the expression for the second
derivative of the phase
\begin{equation} \label{7-7}
\frac{\prt^2\varphi}{\prt\eta^2}=8\frac{\cos{\mu}}{\widetilde{k}^3}[(M^2-2)\widetilde{k}^2+6(M^2-1)].
\end{equation}
As a result, the expression for the condensate density (\ref{2-17}) takes
the form
\begin{equation} \label{2-20}
\delta\rho=V_0\rho_0\sqrt{\frac{2\widetilde{k}}{\pi
r}}\frac{[(M^2-2)\widetilde{k}^2+4(M^2-1)]^{1/4}}{[(M^2-2)\widetilde{k}^2+6(M^2-1)]^{1/2}}
\cos\left(c_s\widetilde{k}r\cos{\mu}-\frac{\pi}4\right),
\end{equation}
where
\begin{equation}\label{tk}
    \widetilde{k}=2\sqrt{M^2\cos^2{\eta}-1}.
\end{equation}
As we see from Eq.~(\ref{2-20}), the linear waves
exist only in the region
\begin{equation}\label{2-21}
    -\arccos(1/M)\leq\eta\leq\arccos(1/M)
\end{equation}
outside the Mach cone.

With the help of Eqs.~(\ref{2-19}) one can find the shape of the
lines of constant phase (e.g. wave crests) $\Phi=kr\cos\mu$ in a
parametric form
\begin{equation}\label{2-20a}
    \begin{split}
    &x=r\cos\chi=\frac{4\Phi}{c_s\widetilde{k}^3}\cos\eta(1-M^2\cos2\eta),\\
    &y=r\sin\chi=\frac{4\Phi}{c_s\widetilde{k}^3}\sin\eta(2M^2\cos^2\eta-1).
    \end{split}
\end{equation}
Small values of $\eta$ correspond to waves in front of the
obstacle. In this case we have
\begin{equation}\label{2-21a}
     \begin{split}
    &x\cong -\frac{\Phi}{2c_s\sqrt{M^2-1}}+\frac{(2M^2-1)\Phi}{4c_s(M^2-1)^{3/2}}\eta^2,\\
    &y\cong \frac{(2M^2-1)\Phi}{2c_s(M^2-1)^{3/2}}\eta,
    \end{split}
\end{equation}
that is the lines of stationary phase take parabolic form
\begin{equation}\label{2-22}
    x(y)\cong -\frac{\Phi}{2c_s\sqrt{M^2-1}}+\frac{c_s(M^2-1)^{3/2}}{(2M^2-1)\Phi}y^2.
\end{equation}
The limiting values $\eta=\pm\arccos{(1/M)}$ correspond to the
lines
\begin{equation}\label{2-23}
    \frac{x}y=\pm\sqrt{M^2-1},
\end{equation}
i.e. far from the obstacle the lines approach to the straight lines
parallel to those forming the Mach cone~(\ref{mach}).
Predictions of the analytical theory are compared with the numerically
calculated wave pattern in Fig.~5 and excellent agreement is found.
\begin{figure}[bt]
\begin{center}
\includegraphics[width=8cm,height=7cm,clip]{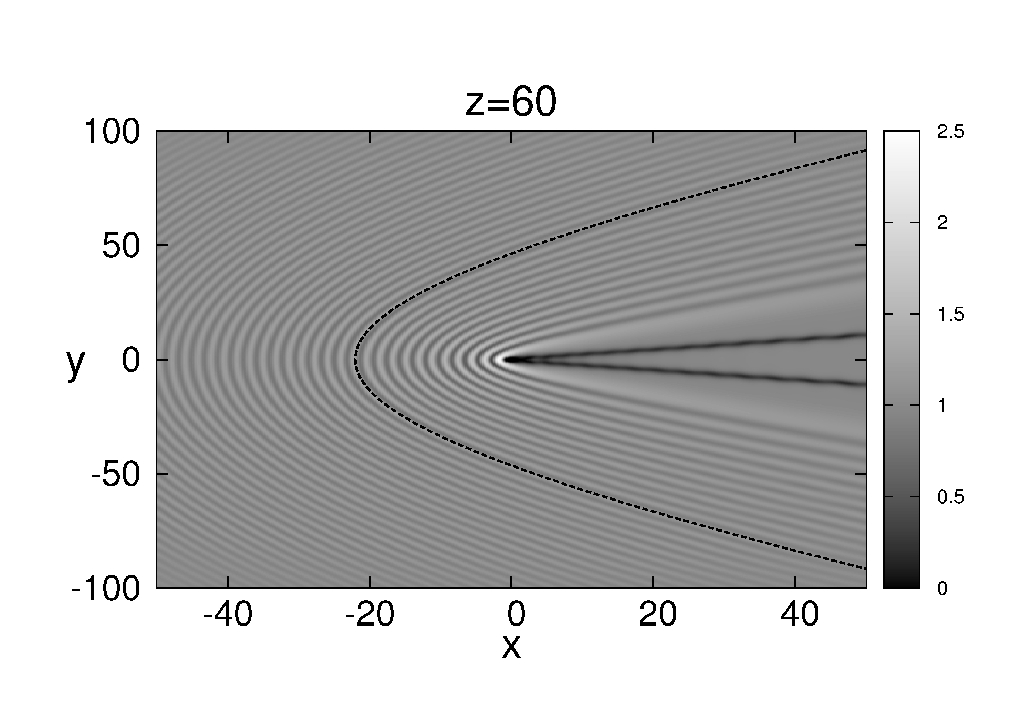}
\caption{Numerically calculated wave pattern corresponding to
diffraction of a light beam on the obstacle embedded into a
photorefractive medium. The plot corresponds to $\ga=0.2$, $U=2$,
and the radius of the reflecting wire to $r=1$. Dashed line
corresponds to linear analytical theory, Eq.~(\ref{2-20a}), for the
line of constant phase; it is shifted to the left to two units of
length from the center of the obstacle due to its finite size in
numerical simulations and better fitting to numerics. }
\end{center}\label{fig5}
\end{figure}

In the region in front of the obstacle where $y=0,\,x<0$, the
perturbations of the light intensity take the simplest form. Here
we have
\begin{equation}\label{9-1}
    k=2c_s\sqrt{M^2-1},
\end{equation}
i.e. the wave length $\lambda=2\pi/k$ is constant and
\begin{equation}\label{9-2}
    \delta\rho=2V_0\rho_0\sqrt{\frac{(M^2-1)^{1/2}}{\pi(2M^2+1)|x|}}
    \cos\left(-2c_s\sqrt{M^2-1}\,x-\frac{\pi}4\right),\quad y=0,\quad x<0.
\end{equation}
The plot illustrating this dependence is shown in Fig.~6. As we see,
approximate formula Eq.~(\ref{2-20}) is accurate enough almost
everywhere except the small vicinity of the obstacle.
\begin{figure}[bt]
\begin{center}
\includegraphics[width=8cm,height=7cm,clip]{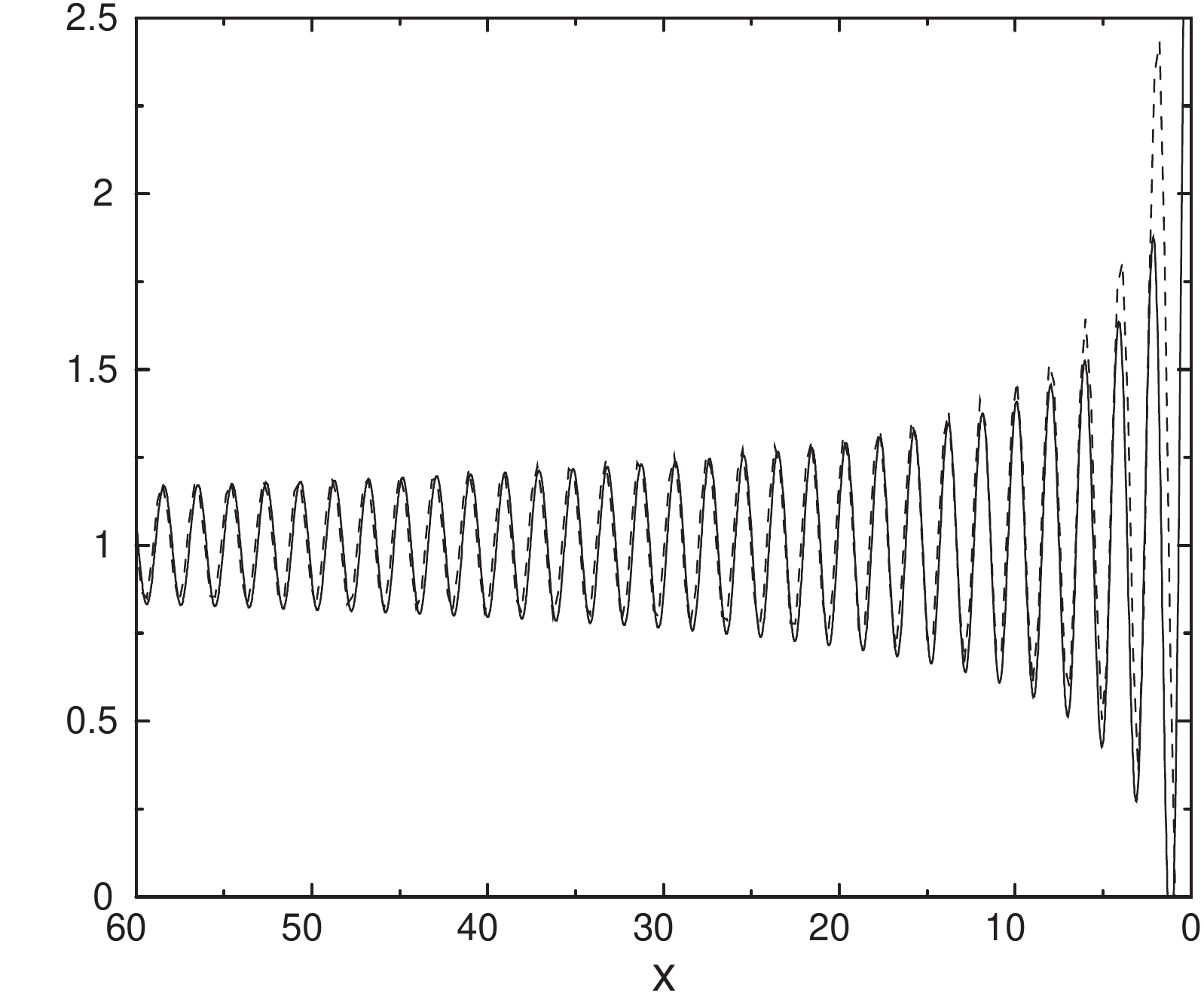}
\caption{Profile of intensity in front of the the obstacle for
$x<0$, $y=0$ and choice of the parameters $\ga=0.2$, $U=2$, $V_0=2.6$. Solid
line corresponds to Eq.~(\ref{9-2}) and dashed line to numerical
solution of Eqs.~(\ref{1-5a},\ref{1-5b}).}
\end{center}\label{fig6}
\end{figure}

As was indicated above, the method of stationary phase used for the
derivation of (\ref{2-20})requires the condition
$|k(\eta)\cos(\chi+\eta)|\gg1$. According to (\ref{2-13})  we have
$k\rightarrow0$ at the Mach cone and the necessary condition is not
fulfilled. To find a wave pattern near the Mach cone one should
return to the investigation of the integral (\ref{2-9}) and
introduce new coordinates along the Mach cone ($\xi$) and normal to
it ($\tau$) (i.e., they are rotated to the angle $\theta$ around the
origin):
\begin{equation}\label{2-24}
x=\xi\cos{\theta}-\tau\sin{\theta},\quad
y=\xi\sin{\theta}+\tau\cos{\theta}.
\end{equation}
In new coordinates equation (\ref{2-9}) takes the form
\begin{equation} {\label{2-26}}
\delta\rho=V_0\rho_0\int\!\!\int\frac{k^2  e^{i(k_{\xi}\xi+k_{\tau}\tau)}}
{(k_{\xi}U\cos{\theta}-k_{\tau}U\sin{\theta})^2-k^2(c_s^2+{k^2}/{4})+i0}\frac{d
k_{\xi}d k_{\tau}}{(2\pi)^2}.
\end{equation}

Far from the obstacle, near the Mach cone, the dependence of the
wave pattern on the $\xi$-coordinate is much slower than dependence
on the $\tau$-coordinate; besides that one has $|k|\ll1$ here. Main
contribution into the integral over $k_{\xi}$ is due to the pole
which position is determined by the equations
\begin{equation}\label{2-27}
(k_{\xi}U\cos{\theta}-k_{\tau}U\sin{\theta})^2-k^2(1+{k^2}/{4})=0, \quad
k_{\xi}^2+k_{\tau}^2=k^2.
\end{equation}
Their approximate solution for $k_{\xi}\ll k_{\tau}\ll1$ is given by
\begin{equation}\label{2-28}
    k_{\xi}=-\frac{k_{\tau}^3}{8\sqrt{M^2-1}}
\end{equation}
where we have taken into account Eq.~(\ref{sound2}). Integration over $k_{\xi}$
yields
\begin{equation}\label{2-29}
    \delta\rho=\frac{V_0\rho_0}{2\sqrt{M^2-1}}\frac{\prt}{\prt\tau}\left[
    \frac1{\pi}\int_0^{\infty}\cos\left(\frac{k_{\tau}^3\xi}{8\sqrt{M^2-1}}-k_{\tau}\tau
    \right)dk_{\tau}\right],
\end{equation}
and with account of the integral representation of the Airy function
\begin{equation}\label{2-30}
    \mathrm{Ai}(z)=\frac1{\pi}\int_0^{\infty}\cos\left(\tfrac13\kappa+z\kappa\right)d\kappa
\end{equation}
we obtain the following expression for the density oscillations in
the vicinity of the Mach cone:
\begin{equation}\label{2-32}
    \delta\rho=-\frac{2V_0\rho_0}{(M^2-1)^{1/6}(3\xi)^{2/3}}\mathrm{Ai}'\left[-\frac{2(M^2-1)^{1/6}}
    {(3\xi)^{1/3}}\tau\right],
\end{equation}
where $\mathrm{Ai}'$ denotes the first derivative of the Airy
function with respect to its argument. Returning to $x$ and $y$
coordinates, we get
\begin{equation}\label{2-32a}
    \delta\rho=-\frac{2V_0\rho_0}{(M^2-1)^{1/6}[3(x\cos\theta+y\sin\theta)]^{2/3}}\mathrm{Ai}'
    \left[-\frac{2(M^2-1)^{1/6}}
    {[3(x\cos\theta+y\sin\theta)]^{1/3}}(-x\sin\theta+y\cos\theta)\right],
\end{equation}
where $\sin\theta=1/M$, $\cos\theta=\sqrt{M^2-1}/M$.

The above formulae allow one to derive expressions for the
dependence of intensity on $y$ coordinate for fixed value of $x$
which may be convenient for comparison with the experiment and
numerical simulations. Far enough from the Mach cone when
Eq.~(\ref{2-20}) can be applied we find dependence of $\chi$ on $y$
from the equation
\begin{equation}\label{2-33}
    \frac{y}x=\tan\eta=\frac{(1+\widetilde{k}^2/2)\tan\eta}{M^2-(1+\widetilde{k}^2/2)},
\end{equation}
then
\begin{equation}\label{2-34}
    r(\eta)=\frac{y}{\sin\chi(\eta)}\, ,
\end{equation}
where $\widetilde{k}(\eta)$ and $\mu(\eta)$ are defined by Eqs.~(\ref{tk}) and (\ref{7-6}).
In the limit $y\gg x$ we have $\chi\to\pi/2$, hence denominator in the rhs of Eq.~(\ref{2-33})
vanishes and
\begin{equation}\label{2-35}
    \widetilde{k}\cong\sqrt{2(M^2-1)}\quad \mathrm{for}\quad y\gg x.
\end{equation}
Comparison with Eq.~(\ref{tk}) gives the limiting value of $\eta$,
\begin{equation}\label{2-36}
    \cos\eta\cong\frac{\sqrt{M^2+1}}{\sqrt{2}M}.
\end{equation}
Substitution of these values of the parameters into Eq.~(\ref{2-20}) yields
\begin{equation}\label{2-37}
    \delta\rho(y)\cong -V_0\rho_0\sqrt{\frac{2M}{\pi(M^2-1)y}}\cos\left[\left(M-1/M\right)y-\pi/4\right]
    \quad \mathrm{for}\quad y\gg x.
\end{equation}
The profile of the wave in the vicinity of the Mach cone is shown in
Fig.~7. As we see, Eq.~(\ref{2-20}) reproduces  the density profile
very well almost everywhere except for a closest vicinity of the
Mach cone and inside it where the density perturbation decays
exponentially according to the behavior of the Airy function in
Eq.~(\ref{2-32a}).

\begin{figure}[bt]
\begin{center}
\includegraphics[width=8cm,height=7cm,clip]{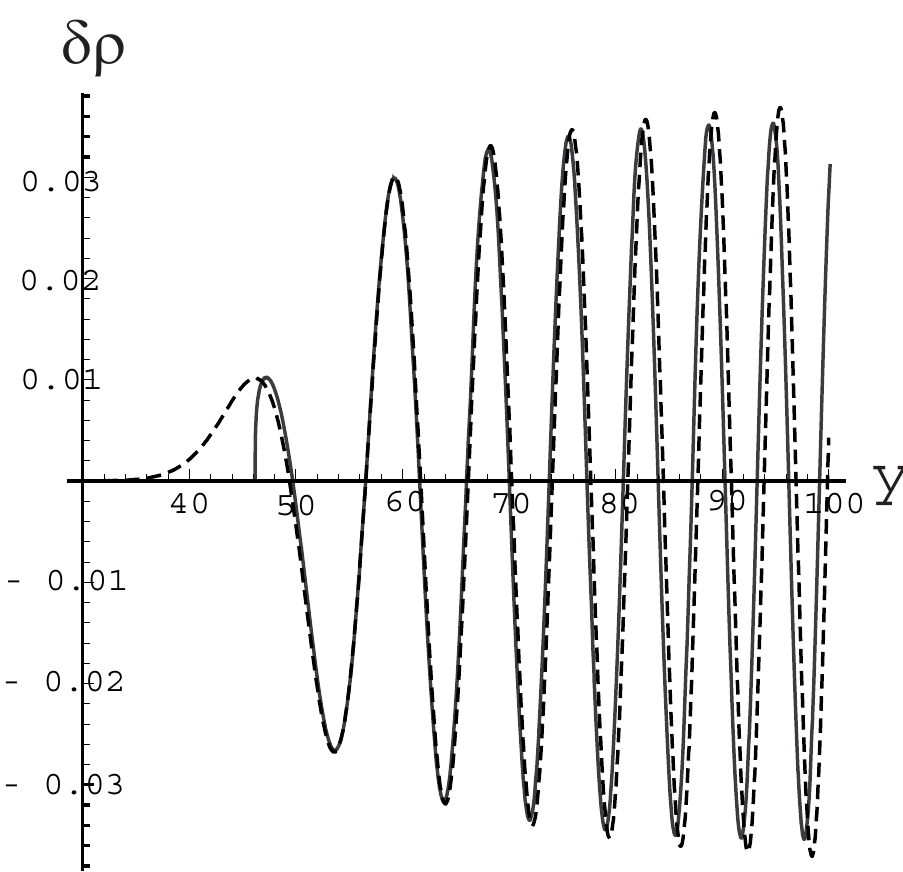}
\caption{Wave pattern near the Mach cone. Solid line corresponds to Eq.~(\ref{2-20})
and dashed line to Eq.~(\ref{2-32a}).}
\end{center}\label{fig7}
\end{figure}

\section{Oblique dark soliton}

Far enough from the obstacle where vorticity is equal to zero and
the light ``flow'' can be considered as potential, we can use the
hydrodynamic representation of equations of light beam evolution.
Here the potential of the obstacle can be neglected (in case of a
reflecting wire it obviously vanishes beyond the surface of the
wire, that is the obstacle is represented by an infinite cylindrical
barrier) and for large enough $z$ the soliton is close to its
stationary state. The profiles of intensity $\rho$ and
``velocities'' $u,\,v$ can be found analytically as a solution of
stationary equations
\begin{equation}\label{3-1}
    (\rho u)_x+(\rho v)_y=0,
\end{equation}
and
\begin{equation}\label{3-2}
\begin{split}
      uu_x+vu_y+\left(\frac{\rho}{1+\ga\rho}\right)_x+\left(\frac{\rho_x^2+\rho_y^2}{8\rho^2}-
   \frac{\rho_{xx}+\rho_{yy}}{4\rho}\right)_x=0,\\
   uv_x+vv_y+\left(\frac{\rho}{1+\ga\rho}\right)_y+\left(\frac{\rho_x^2+\rho_y^2}{8\rho^2}-
   \frac{\rho_{xx}+\rho_{yy}}{4\rho}\right)_y=0,
   \end{split}
\end{equation}
with boundary conditions (in this Section we assume $\rho_0=1$)
\begin{equation}\label{3-3}
   \rho=1,\quad u=U,\quad v=0\quad\text{at}\quad |x|\to\infty.
\end{equation}
To simplify calculations, it is convenient to notice that one of
equations (\ref{3-2}) can be replaced by the condition of zero
vorticity
\begin{equation}\label{3-4}
    u_y-v_x=0
\end{equation}
which is fulfilled for the potential flow in the soliton solution.

We look for the solution in the form
\begin{equation}\label{3-5}
    \rho=\rho(\theta),\quad u=u(\theta),\quad v=v(\theta),\quad
    \text{where}\quad \theta=x-ay.
\end{equation}
The parameter $a$ determines a slope of the oblique soliton in the
$x,y$ plane. Then equations (\ref{3-1}) and (\ref{3-4}) with account
of conditions (\ref{3-3}) give after simple calculation the
expressions for the components of the ``flow velocity'' in terms of
the light intensity
\begin{equation}\label{3-6}
    u=\frac{U(1+a^2\rho)}{(1+a^2)\rho},\quad v=-\frac{aU(1-\rho)}{(1+a^2)\rho}.
\end{equation}
Substitution of these expressions into any equation (\ref{3-2}) and integration
of the resulting equation yields
\begin{equation}\label{3-7}
    \frac18(1+a^2)^2(\rho'^2-2\rho\rho^{\prime\prime})+(1+a^2)\frac{\rho^3}{1+\ga\rho}
    -\left(\frac{U^2}2+\frac{1+a^2}{1+\ga}\right)\rho^2+\frac{U^2}2=0
\end{equation}
where an integration constant is chosen in accordance with the conditions (\ref{3-3}).
This equation can be integrated once more to give
\begin{equation}\label{3-8}
    \frac{(1+a^2)^2}8\left(\frac{d\rho}{d\theta}\right)^2=
    -\frac{(1+a^2)\rho}{\ga^2}\ln(1+\ga\rho)+\left(
    \frac{1+a^2}{(1+\ga)\ga}-\frac{U^2}2\right)\rho^2+
    \left(U^2+\frac{1+a^2}{\ga^2}\ln(1+\ga)-\frac{1+a^2}{\ga(1+\ga)}\right)\rho-\frac{U^2}2
\end{equation}
where (\ref{3-3}) is also taken into account. For a given ``Mach
number'' $M=(1+\ga)U$ the soliton solution depends on the slope
parameter $a$ alone.

Now we notice that expressions for the flow velocity field
(\ref{3-6}) in terms of intensity $\rho$ do not depend on the
nonlinear properties of the medium but are determined completely by
the ``continuity'' equation and the condition (\ref{3-4}) of
potentiality of the flow. Therefore we can change the ``reference
frame'' in such a way that the transversal velocity (wave vector
$\bu$) is equal to zero at $|\theta|\to\infty$. This means that we
rotate the reference frame to the angle $\phi=\arctan a$ and pass to
the frame ``moving'' with ``velocity'' $(U\cos\phi,U\sin\phi)$ as
$z$ increases, which means the change of coordinates
\begin{equation}\label{4-1}
    \begin{split}
    \widetilde{x}=x\cos\phi-y\sin\phi-U\cos\phi\cdot z,\\
    \widetilde{y}=x\sin\phi+y\cos\phi-U\sin\phi\cdot z.
    \end{split}
\end{equation}
Correspondingly, the ``velocity'' field transforms as
\begin{equation}\label{4-2}
    \begin{split}
    \widetilde{u}=(u-U)\cos\phi-v\sin\phi,\\
    \widetilde{v}=(u-U)\sin\phi+v\cos\phi.
    \end{split}
\end{equation}
Substitution of (\ref{3-6}) gives
\begin{equation}\label{4-3}
    \widetilde{u}=c\left(\frac1{\rho}-1\right),\quad
    \widetilde{v}=0\, ,
\end{equation}
where we have introduced the parameter
\begin{equation}\label{4-4}
    c=\frac{U}{\sqrt{1+a^2}}.
\end{equation}
In new variables the velocity field does not have a component along $\widetilde{y}$ coordinate.
The variable $\theta$ takes the form $\theta=\sqrt{1+a^2}(\widetilde{x}+cz)$ and hence the
intensity $\rho$ does not depend on $\widetilde{y}$ coordinate. Thus, in new coordinate
system we have a 1D dark soliton moving with velocity $c$ in negative direction of $\widetilde{x}$
axis. This transformation will be used below in the study of stability of dark solitons.

Introduction of the parameter $c$ permits one to represent equation (\ref{3-8}) as
\begin{equation}\label{4-5}
    \frac{1}{8}\left(\frac{d\rho}{d\xi}\right)^2=-\frac{\rho}{\ga^2}\ln(1+\ga\rho)+\left(
    \frac{1}{(1+\ga)\ga}-\frac{c^2}2\right)\rho^2+
    \left(c^2+\frac{1}{\ga^2}\ln(1+\ga)-\frac{1}{\ga(1+\ga)}\right)\rho-\frac{c^2}2\equiv Q(\rho),
\end{equation}
where $\xi=\widetilde{x}+cz$. The function $Q(\rho)$ has a double
zero at $\rho=1$ which corresponds to the tails of soliton. Another
zero at $\rho=\rho_m$ corresponds to the minimal intensity at the
center of soliton, which is, therefore, related to the parameter $c$
as
\begin{equation}\label{4-6}
    c=\frac1{1-\rho_m}\left[\frac{2\rho_m}{\ga}\left(\frac1{\ga}\ln\frac{1+\ga}{1+\ga\rho_m}-
    \frac{1-\rho_m}{1+\ga}\right)\right]^{1/2}.
\end{equation}
Taking into account Eq.~(\ref{4-4}) we find expression for the slope $a$ as a function of $\rho_m$:
\begin{equation}\label{4-7}
    a=\left[\frac{U^2(1-\rho_m)^2\ga}{2\rho_m\left(\frac1{\ga}\ln\frac{1+\ga}{1+\ga\rho_m}
    -\frac{1-\rho_m}{1+\ga}\right)}-1\right]^{1/2}.
\end{equation}
The slope of the most shallow solitons with $\rho_m\to1$ is equal to
\begin{equation}\label{4-8}
    a_{min}=\sqrt{(1+\ga)^2U^2-1}=\sqrt{M^2-1}
\end{equation}
that is it coincides with the Mach cone.

The profile of the light intensity across the oblique soliton can be
obtained by  a straightforward numerical integration of
Eq.~(\ref{4-5}). In Fig.~8 we compare such a profile with the
profile of the diffraction pattern obtained by  direct numerical
simulation using original Eqs.~(\ref{1-5a},\ref{1-5b}). Good
agreement between these two profiles confirms that the pattern in
Fig.~2 inside the Mach cone indeed consists of oblique dark solitons
generated by nonlinear diffraction of the light beam on the
obstacle.
\begin{figure}[bt]
\begin{center}
\includegraphics[width=8cm,height=6.5cm,clip]{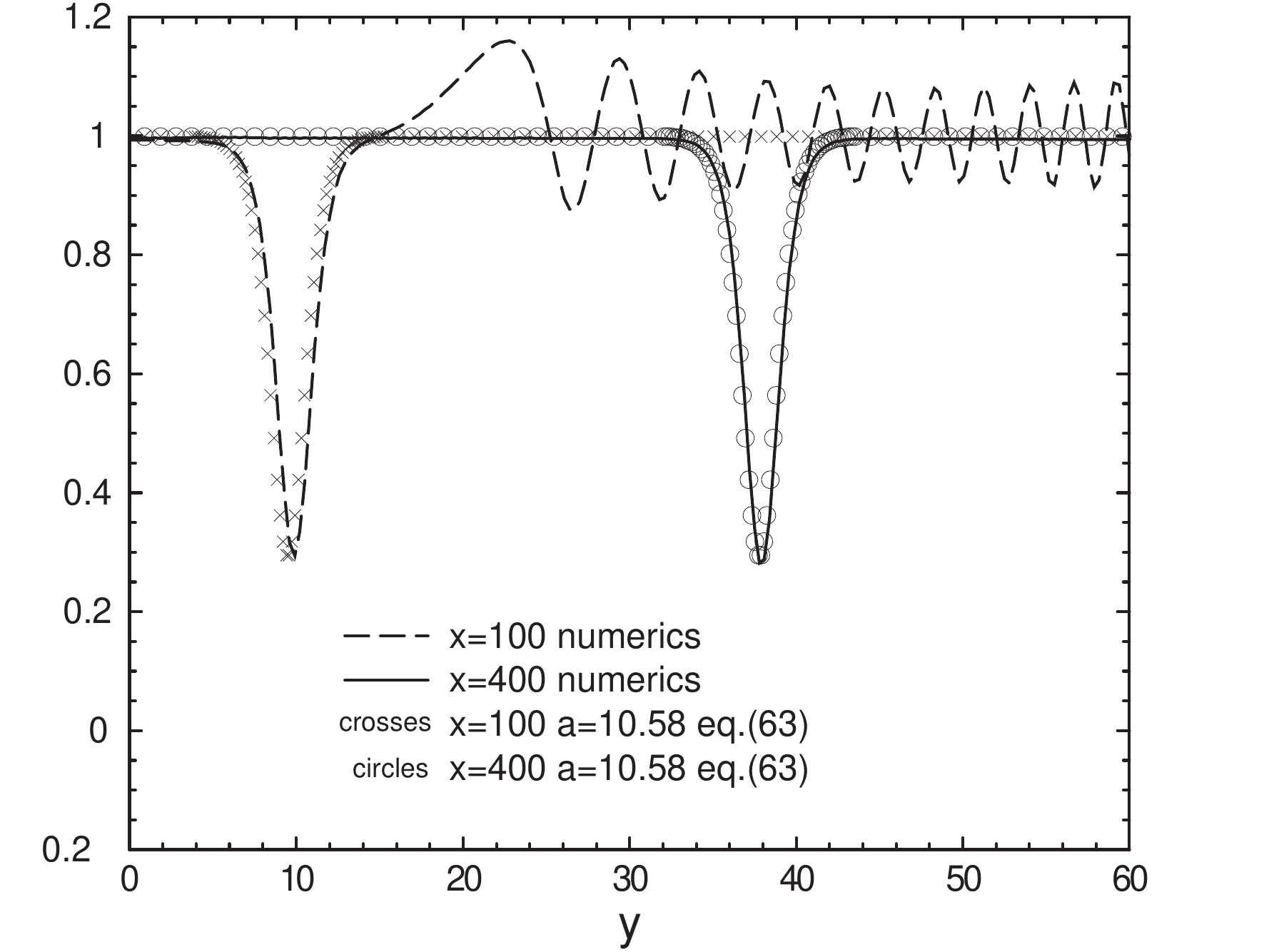}
\caption{Profiles of the intensity distributions for $x=100$ (dashed
line), $x=400$ (solid line) and $y>0$ obtained from numerical
solution of the equation (\ref{1-5a}) with the nonlinear term given
by (\ref{1-5b}). These profiles are compared with the soliton
profiles obtained by solutions of Eq.~(\ref{3-8}) with slope
$a=10.58$ shown as functions of $y$ at the same values of $x$
($x=100$ corresponds to ``crosses'' and $x=400$ to ``circles'').}
\end{center}\label{soliton}
\end{figure}

\section{Stability of oblique solitons}

The solitons profiles investigated in the preceding Section are
reached asymptotically as $z\to\infty$. However, the pattern
calculated for finite $z$ and shown in Fig.~2 indicate that some
oscillations of intensity take place along the oblique solitons.
Amplitude of these oscillations increases with distance from the
obstacle what leads to generation of vortices at the end points of
solitons. In fact, instability of dark 2D solitons with respect to
transverse perturbations is well known as well as development of
this instability to formation of vortices (see \cite{ka03}). But in
the case of formation of dark solitons in the flow of Bose-Einstein
condensate past an obstacle it was found \cite{egk06} that the
amplitude of oscillations decreases with growth of time at fixed
distance from the obstacle for large enough value of the oncoming
flow velocity. This suggests that absolute instability of dark
solitons transforms into their convective instability in the
reference frame attached to the obstacle at some critical value of
the flow velocity \cite{kp07}. This means that wave packets built of
unstable modes of soliton's disturbance are convected so fast by the
flow that they cannot develop at finite distance from the obstacle.
The criterion of transition to the convective instability for
Bose-Einstein condensate evolving according to the Gross-Pitaevskii
equation (see (\ref{1-6})) was derived in \cite{kp07} and here we
shall extend the analysis of \cite{kp07} to the photorefractive
equation (\ref{1-5}).

\subsection{Shallow solitons (Kadomtsev-Petviashvili approximation)}

The theory is especially simple in the limit of small-amplitude
solitons when the GNLS equation (\ref{1-5a}) can be reduced to the
Kadomtsev-Petviashvili (KP) equation by means of standard reductive
perturbation theory, which yields
\begin{equation}\label{kad-pet}
    \left[-2c_s\rho'_z+2c_s^2\rho'_\tx+\left(3f'(\rho_0)+\rho_0f^{\prime\prime}(\rho_0)\right)\rho'\rho'_\tx-
    \tfrac14\rho'_{\tx\tx\tx}\right]_\tx+c_s^2\rho'_{\ty\ty}=0,
\end{equation}
where $\rho'\ll\rho_0$ denotes the intensity perturbation small
compared with the background intensity $\rho_0$, $\tx$ is a
coordinate along a soliton and $\ty$ is a transverse coordinate. We
transform it to the standard form by introducing the new variables
\begin{equation}\label{var1}
    \tilde{z}=\frac{z}{2c_s},\quad \tilde{\xi}=\tx + c_s z,\quad \tilde{\eta}=\frac{\ty}{c_s},\quad
    \tro=-\tfrac13(3f'(\rho_0)+\rho_0f^{\prime\prime}(\rho_0))\rho'
\end{equation}
to obtain
\begin{equation}\label{kp2}
    \left(\tro_z+3\tro\tro_{\txi}+\tfrac14\tro_{\txi\txi\txi}\right)_{\txi}=\tro_{\teta\teta}.
\end{equation}
As is well known, the KP equation (\ref{kp2}) has the soliton
solution
\begin{equation}\label{kdv}
    \tro_s=\frac{s}{\cosh^2[\sqrt{s}(\txi-s\tz)]}=\frac{s}{\cosh^2[\sqrt{s}(\tx+(c_s-s/(2c_s))z)]}
\end{equation}
where the parameter $s$ is small,
\begin{equation}\label{s}
    \frac{s}{c_s^2}\ll1,
\end{equation}
in accordance with the condition that the soliton is shallow. This
solution solution is written in the reference frame with $\bu\to 0$
as $x\to\infty$.

The soliton solution (\ref{kdv}) is unstable with respect to transverse perturbations
\cite{kp-1970,zakharov-1975}. If we perturb the solution (\ref{kdv}) along $\ty$ axis,
\begin{equation}\label{kp3-3}
    \tro=\tro_s(\zeta)+\delta\tro,\quad \delta \tro=W(\zeta)\exp(\Gamma z+ip\ty),\quad \zeta=\tx+c\ty,
\end{equation}
then in linear approximation we obtain equation for $W$:
\begin{equation}\label{kp3-4}
    \left[-W_{\zeta\zeta\zeta}+4sW_\zeta-12(\tro_sW)_\zeta\right]_\zeta-4p^2W=
    4\Gamma W_\zeta.
\end{equation}
This eigenvalue problem was studied in \cite{zakharov-1975,aps-1997}
where the following spectrum for the instability growth rate was
obtained
\begin{equation}\label{kp3-5}
    \Gamma(p)=({p}/{\sqrt{3}})\sqrt{s-{2p}/{\sqrt{3}}}.
\end{equation}
Thus, in the reference frame with $\bu\to 0$ at $x\to\infty$ the soliton is absolutely unstable.

However, we are interested in the behavior of the soliton
transformed to the reference frame ``attached'' to the obstacle by
substitution (\ref{4-1}):
\begin{equation}\label{kdv2}
    \tro_s={s}\cosh^{-2}\left\{\sqrt{\frac{s}{1+a^2}}\left[x-ay+\left(\left(c_s-\frac{s}{2c_s}\right)
    \sqrt{1+a^2}-U\right)z\right]\right\}.
\end{equation}
The relationship between the soliton parameter $s$ and the slope $a$
follows from the condition that the oblique soliton solution does
not depend on $z$:
\begin{equation}\label{s2}
    s=c_s^2\left(1-\frac{M^2}{1+a^2}\right),
\end{equation}
where we took into account (\ref{sound2}) and (\ref{s}). After the
transformation to the ``obstacle'' frame we easily get the
dispersion relation
\begin{equation}\label{kp3-7}
%\begin{split}
    \om=\om(p)=\mu p+i\frac{p}{\sqrt{3}}\sqrt{s-\frac{2p}{\sqrt{3}}},\quad
    \mu=U\sin\theta=\frac{Ma}{(1+\ga)\sqrt{1+a^2}}
%   \end{split}
\end{equation}
for waves propagating along oblique soliton with the wave number
$p$. The stability of the soliton is determined by the asymptotic
behavior of the wave packets built from harmonic waves. Due to the
term $\mu p$ in the dispersion relation, the wave packets are
convected by the flow along the soliton. If they are convected fast
enough, then amplitude of the unstable disturbance cannot increase
at fixed distance from the obstacle and,  as a result, the soliton
is just convectively unstable \cite{kp07}. As was shown in
\cite{kp07}, for shallow KP solitons the criterion of transition
from absolute to convective instability reads
\begin{equation}\label{crit1}
    \mu^2>s.
\end{equation}
Then substitution of Eqs.~(\ref{s2}) for $s$ and (\ref{kp3-7}) for $\mu$ gives at once
\begin{equation}\label{crit2}
    M>1.
\end{equation}
Thus, the shallow solitons are convectively unstable for
``supersonic'' values of transverse wave vector $U$.

\subsection{Deep solitons}

Now we consider stability of soliton solutions of the photorefractive equation (\ref{1-5})
with dropped external potential:
\begin{equation}\label{d1}
    i\frac{\prt\psi}{\prt z}+\frac1{2}\Delta_\bot\psi-
    \frac{|\psi|^2}{1+\gamma|\psi|^2}\psi =0.
\end{equation}
Stability of solitons for the case of 2D NLS equation ($\ga=0$) was studied in \cite{kt-1988}.
We shall write the soliton solution of Eq.~(\ref{d1}) in the form
\begin{equation}\label{d2}
    \psi_s(\zeta)=\sqrt{\rho_s(\zeta)}\exp\left(i\phi_s(\zeta)-\frac{iz}{1+\ga}\right)
\end{equation}
where $\rho_s(\zeta)$ is given by the solution of Eq.~(\ref{4-5}) and
\begin{equation}\label{d3}
    \frac{\prt\phi_s}{\prt\zeta}=c\left(\frac1{\rho_s(\zeta)}-1\right).
\end{equation}
The disturbed function $\psi$ can be taken in the form
\begin{equation}\label{d4}
    \psi=\psi_s(\zeta)+(\psi'+i\psi^{\prime\prime})\exp\left(i\phi_s(\zeta)-\frac{iz}{1+\ga}\right).
\end{equation}
Here $\psi'$ and $\psd$ depend on $y$ and $z$ as $\exp(ipy+\Gamma z)$
Substitution of Eq.~(\ref{d4}) into (\ref{d1}) and linearization with respect to $\psi'$ and $\psd$
yields the linear spectral problem
\begin{equation}\label{d5}
\left(
\begin{array}{cc}
-A & L_1 \\ L_2 & A
\end{array}
\right)\left(
\begin{array}{c}
\psi^{\prime\prime} \\ \psi'
\end{array}\right) = \Gamma
\left(
\begin{array}{c}
\psi^{\prime\prime} \\ -\psi'
\end{array}\right),
\end{equation}
where
\begin{equation}
A=\frac{c}{\rho_s}\left(\frac{\partial}{\partial
\xi}-\frac{\rho_{s,\xi}}{2\rho_s}\right),
\end{equation}
\begin{equation}
L_1=\frac{1}2\frac{\partial^2}{\partial\xi^2}+\frac{1}2(c^2-p^2)+\frac{1}{1+\gamma}
-\frac{1}2\frac{c^2}{\rho_s^2} -\frac{3\rho_s+\ga\rho_s^2}{(1+\ga\rho_s)^2},
\end{equation}
\begin{equation}
L_2=\frac{1}2\frac{\partial^2}{\partial\xi^2}+\frac{1}2(c^2-p^2)+\frac{1}{1+\gamma}
-\frac{1}2\frac{c^2}{\rho_s^2}-\frac{\rho_s}{1+\ga\rho_s}.
\end{equation}
The function $\rho_s$ is considered here as known for a given value
of the soliton velocity $c$; hence the system (\ref{d5}) can be
solved numerically which yields the spectrum of the growth rate
$\Gamma=\Gamma(p)$ for all values of $c$.

Again, we transform this solution to the reference frame attached to
the obstacle and arrive at the dispersion relation
\begin{equation}\label{d6}
    \om(p)=\mu p +i\Gamma(p).
\end{equation}
This equation determines implicitly the function $p=p(\om)$. The
type of stability is determined by the location of branching points
$p_{br}$ of this function (see, e.g., \cite{ll-10}) where
$d\om/dp=0$ what gives the equation
\begin{equation}\label{d7}
    \mu=-i\frac{d\Gamma}{dp} \, ,
\end{equation}
which determines the branching point $p_{br}$ as a function of $\mu$
at a given value of $c$.
As was shown in \cite{kp07}, the critical value $\mu_{cr}$ of transition
from absolute instability to convective one
is determined by the condition that the function $p_{br}(\mu)$ has a
branching point at $\mu=\mu_{cr}$. This gives the equation
\begin{equation}\label{d8}
    \left.\frac{d^2\Gamma}{dp^2}\right|_{p=p_{cr}}=0 \, ,
\end{equation}
solution of which gives the critical value $p_{cr}$ for a given $c$.
Example of the plot of the absolute value of the function
$\Gamma(p)$ is shown in Fig.~9 for $\ga=0.1$ and soliton velocity
$c$ corresponding to the minimal intensity $\rho_m=0.2$ and
calculated by means of Eq.~(\ref{4-6}). It has an inflection point
at $p=p_{cr}$ in the region where $\Gamma(p)$ is purely imaginary;
thus, $p_{cr}$ can be calculated for a set of values of $\rho_m$ in
the interval $0\leq\rho_m\leq 1$.
\begin{figure}[bt]
\begin{center}
\includegraphics[width=9cm,height=6cm,clip]{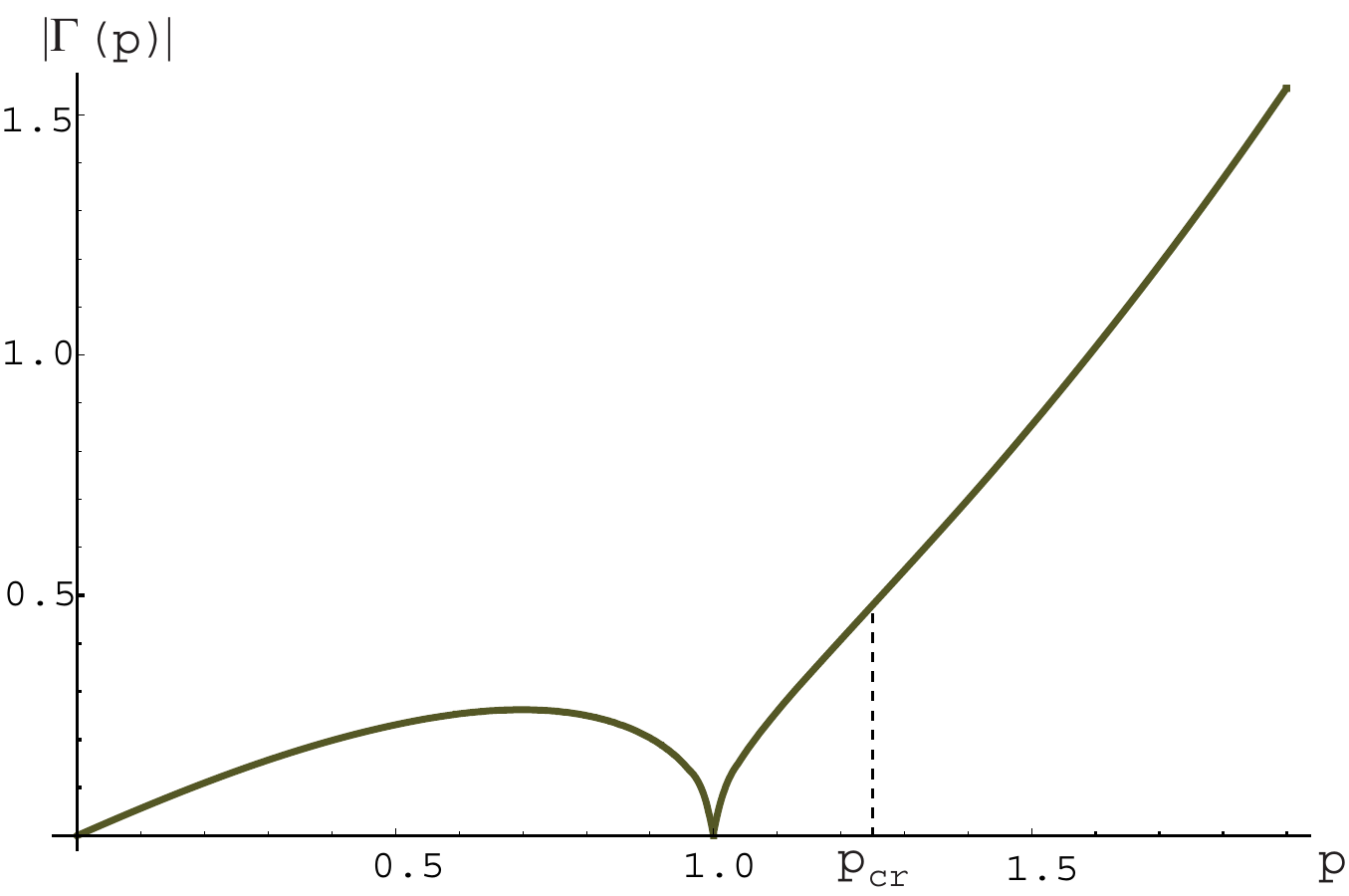}
\caption{Absolute value of the growth rate $\Gamma$ as a function of
the wave number $p$ of a harmonic transverse perturbation for $\ga=1$
and $\rho_m=0.2$.}
\end{center}\label{fig9}
\end{figure}

When $p_{cr}$ is found as a function of $\rho_m$, we can substitute
its value into Eq.~(\ref{d7}) to obtain the critical value of $\mu$,
again, as a function of $\rho_m$:
\begin{equation}\label{d9}
    \mu_{cr}(\rho_m)=-i\left.\frac{d\Gamma(p,\rho_m)}{dp}\right|_{p=p_{cr}}.
\end{equation}
Now we substitute the relation
\begin{equation}\label{d10}
    c=\frac{M}{(1+\ga)\sqrt{1+a^2}}
\end{equation}
into $\mu=Ma/((1+\ga)\sqrt{1+a^2})$ to find $\mu=ca$ which gives the
slope parameter as a function of $\rho_m$:
\begin{equation}\label{d11}
    a(\rho_m)=\frac{\mu_{cr}(\rho_m)}{c(\rho_m)}
\end{equation}
where $c(\rho_m)$ is given by Eq.~(\ref{4-6}).
At last, substitution of this function into Eq.~(\ref{d10}) yields $M_{cr}$
as a function of $\rho_m$:
\begin{equation}\label{d12}
    M_{cr}(\rho_m)=(1+\ga)c(\rho_m)\sqrt{1+a^2(\rho_m)}.
\end{equation}
Equations (\ref{d11}) and (\ref{d12}) determine the critical value
of Mach number as a function of the slope $a$ in a parametric form
with $0\leq\rho_m\leq1$ playing a role of the parameter. Results of
numerical computation of this function for several values of $\ga$
are shown in Fig.~10.
\begin{figure}[bt]
\begin{center}
\includegraphics[width=9cm,height=6cm,clip]{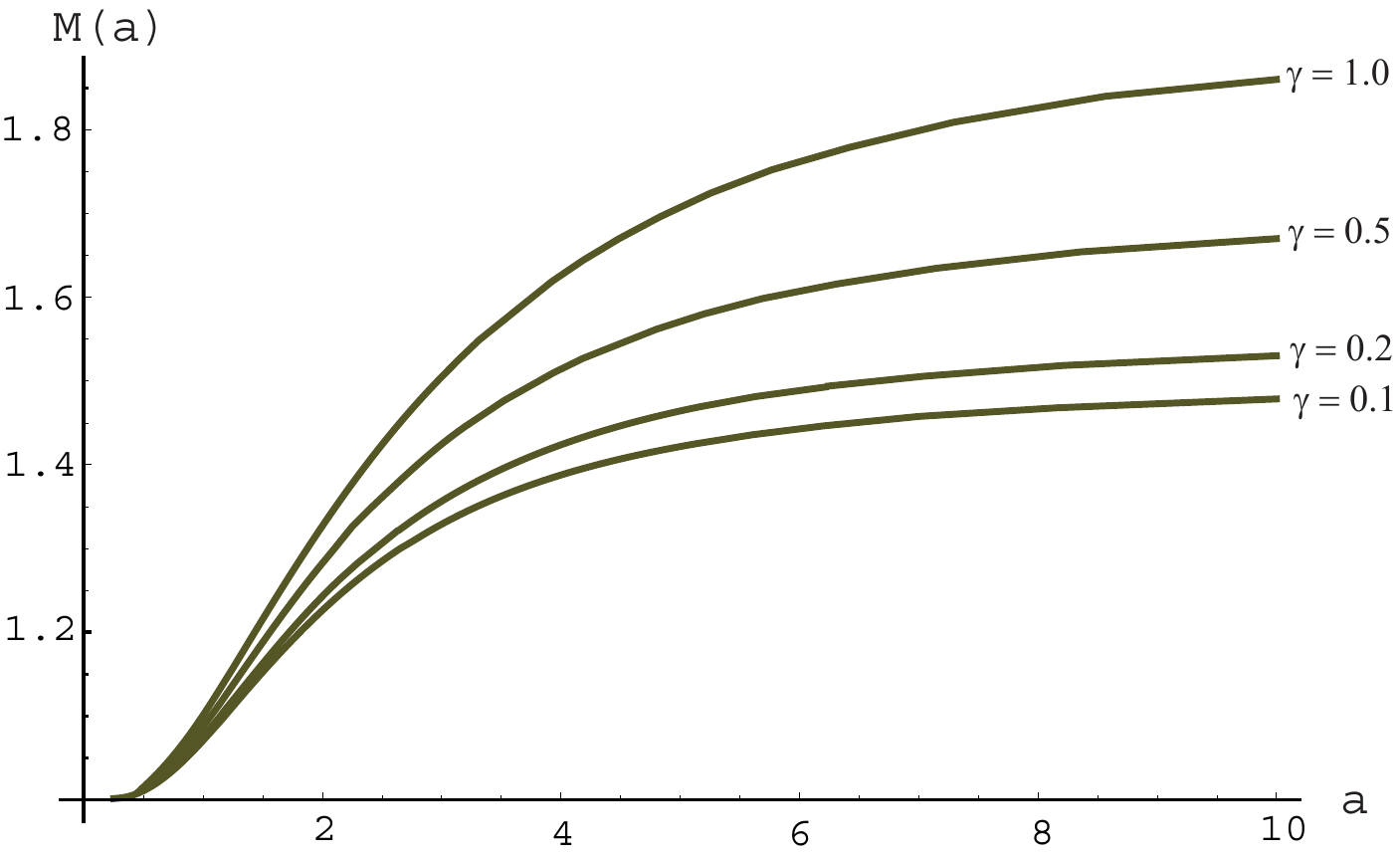}
\caption{Boundary between regions of absolute and convective instabilities
for several values of the saturation parameter $\ga$.}
\end{center}\label{fig10}
\end{figure}
Below these curves oblique solitons are absolutely unstable and
cannot be created by the ``flow'' of light past an obstacle:
perturbation of the ``flow'' behind the obstacle decays into
vortices without formation of solitons. Above these curves oblique
solitons become just convectively unstable and their length grows up
faster than they decay into vortices. Hence, vortices exist at the
end points of solitons only and there is a region where the soliton
profile is close to the stationary solution found in the preceding
Section which was confirmed by numerical simulations.

\section{Conclusion}

In this paper, we have developed the theory of formation of the wave
pattern of light propagating through nonlinear photorefractive
medium with a reflecting wire embedded in the medium. The light beam
is supposed to be tilted with respect to the wire which creates the
``flow of light past an obstacle'' analogous to that realized in
experiments on superfluid flow of Bose-Einstein condensate past an
obstacle. An analogy between propagation of light beams and
superfluid dynamics suggests that diffraction pattern similar to
what was observed and predicted theoretically can be found in
optical experiments. We have shown that the diffraction pattern
consists of two regions separated by the ``Mach cone'' outside which
the so-called ``ship waves'' are located while outside this ``Mach
cone'' the nonlinear dispersive shocks generating oblique
soliton trains are situated. The simplest case when just a
single soliton is generated is studied in detail. The main
parameters of the oblique optical soliton are determined and
it is shown that it is actually stable (more precisely, {\it
convectively unstable}) with respect to small transverse
perturbations for large enough values of transverse wave vector of
the light beam. Detailed theory of ``ship waves'' is also given. All
our findings are confirmed by numerical simulations. Since optical
experiments seem more feasible than the experiments with ultra cold
gases, one may hope that our predictions could be verified
experimentally.

\subsection*{Acknowledgments}

Work of EGK and AG was supported by FAPESP/CNPq (Brazil) and
work of YGG and AMK was supported by RFBR (Russia).


\begin{thebibliography}{99}

\bibitem{ka03} Yu.S. Kivshar and G.P. Agrawal, {\it Optical solitons.
From Fibers to Photonic Crystals,} (Academic Press, Amsterdam, 2003).

\bibitem{ps03} L.P. Pitaevskii and S. Stringari,  Bose-Einstein
Condensation, Cambridge University Press, Cambridge, 2003.

\bibitem{bl54} T.B. Benjamin and M.J. Lighthill,
{ Proc. Roy. Soc.} {\bf A224}, 448 (1954).

\bibitem{sagdeev} R.Z. Sagdeev, Collective processes and shock waves in
rarified plasma, in {\it Problems of Plasma Theory,} M.A. Leontovich, Ed.,
Vol. 5, Atomizdat, Moscow, (1964) (in Russian).

\bibitem{whitham1} G.B. Whitham,
{ Proc. Roy. Soc. London} {\bf A283}, 238 (1965).

\bibitem{GP1} A.V. Gurevich and L.P. Pitaevskii,  {\it Zh. Eksp. Teor. Fiz.} {\bf 65,}
590 (1973) [{ Sov. Phys. JETP} {\bf 38}, 291 (1974)].

\bibitem{FFML} H. Flaschka, M.G. Forest, and D.W. McLaughlin,
{ Commun. Pure Appl. Math.,} {\bf 33,} 739--784 (1980).

\bibitem{DN}
B.A. Dubrovin and S.P. Novikov,  {\it Hydrodynamics of weakly
deformed soliton lattices. Differential geometry and Hamiltonian
theory,} { Russian  Math. Surveys,} {\bf 44,} 35--124 (1989).

\bibitem{tsarev} S.P. Tsarev, {Izv. Akad. Nauk,} {\bf 54,} 1048 (1990);
[{Math. USSR Izvestia,} {\bf 37,} 397 (1991)].

\bibitem{GKE} A.V. Gurevich, A.L. Krylov, and G.A. El,  { Zh. Eksp. Teor. Fiz.} {\bf 101,}
1797 (1992) [Sov. Phys. JETP, {\bf 74} 957-962 (1992)].

\bibitem{kamch2000}  A.M. Kamchatnov, {\it Nonlinear Periodic Waves and
Their Modulations---An Introductory Course}, World Scientific,
Singapore (2000).

\bibitem{el05}
G.A. El,  {Chaos,} {\bf 15}, 037103 (2005).

\bibitem{bec1} M.H. Anderson, J.R. Ensher, M.R. Matthews, C. E.
Wieman, E.A. Cornell, Science {\bf 269}, 198 (1995).

\bibitem{bec2} K.B. Davis, M.-O. Mewes, M.R. Andrews, N.J. van Druten,
D.S. Durfee, D.M. Kurn, and W. Ketterle, Phys. Rev. Lett. { 75} (1995) 3969.

\bibitem{bec3} C.C. Bradley, A. Sackett, and R.G. Hulet, Phys. Rev. Lett. {75}
(1995) 1687; {\it ibid} {79} (1997) 1170(E); {\it ibid} {78} (1997) 985.

\bibitem{damski04} B. Damski, Phys.Rev. A {\bf 69,} 043610 (2004).

\bibitem{kgk04} A.M. Kamchatnov, A. Gammal, and R.A. Kraenkel,
Phys. Rev. A {\bf 69,} 063605 (2004).

\bibitem{simula} T.P. Simula, P. Engels, I. Coddington, V. Schweikhard, E.A. Cornell,
and R.J. Ballagh, Phys. Rev. Lett. {\bf 94,}  080404 (2005).

\bibitem{hoefer}  M.A. Hoefer, M.J. Ablowitz, I. Coddington, E.A. Cornell,
P. Engels, and V. Schweikhard, Phys. Rev. { A} {\bf 74,} 023623 (2006).

\bibitem{engels07} P. Engels and C. Atherton, Phys. Rev. Lett. {\bf 99,} 160405 (2007).

\bibitem{fleischer} W. Wan, S. Jia, and J.W. Fleischer, Nature Physics, {\bf 3,}
46 (2007).

\bibitem{trillo} N. Ghofraniha, C. Conti, G. Ruocco, and S. Trillo,
Phys. Rev. Lett. {\bf 99,} 043903 (2007).

\bibitem{fleischer3} C. Barsi, W. Wan, C. Sun, and J.W. Fleischer,
Optics Lett., {\bf 32,} 2930 (2007).

\bibitem{fleischer2}  S. Jia, W. Wan, and J.W. Fleischer,
Phys. Rev. Lett. {\bf 99,} 223901 (2007).

\bibitem{el07} G.A. El, A. Gammal, E.G. Khamis, R.A. Kraenkel, and A.M. Kamchatnov,
Phys. Rev. A {\bf 76,} 053813 (2007).

\bibitem{LL6} L.D Landau and E.M. Lifshitz, {\it Fluid Mechanics}, Pergamon, Oxford, (1987).

\bibitem{karpman} V.I. Karpman, {\it Nonlinear Waves in Dispersive Media,}
Nauka, Moscow, 1973.

\bibitem{GKKE95} A.V. Gurevich, A.L. Krylov, V.V. Khodorovskii and
G.A. El, { JETP,} {\bf 81,}  87 (1995); {\bf 82,}  709 (1996).

\bibitem{ek06} G.A. El and A.M. Kamchatnov, Phys. Lett A {\bf 350,} 192 (2006);
erratum: Phys. Lett. A {\bf 352,} 554 (2006).

\bibitem{egk06} G.A. El, A. Gammal, and A.M. Kamchatnov,
Phys. Rev. Lett. {\bf 97,} 180405 (2006).

\bibitem{kp07} A.M. Kamchatnov and L.P. Pitaevskii, arxiv: 0712.1891.

\bibitem{caruso} I. Carusotto, S.X. Hu, L.A. Collins, and A. Smerzi,
Phys. Rev. Lett. {\bf 97,} 260403 (2006).

\bibitem{gegk07} Yu.G. Gladush, G.A. El, A. Gammal, and A.M. Kamchatnov,
Phys. Rev. A {\bf 75,} 033619 (2007).

\bibitem{gk07} Yu.G. Gladush and A.M. Kamchatnov,  Zh. Eksp. Teor. Fiz. {\bf 132,} 589 (2007)
[JETP, {\bf 105,} 520 (2007)].

\bibitem{sl92} G.A. Swartzlander, Jr. and C.T. Law, Phys. Rev. Lett. {\bf 69,}
2503 (1992).

\bibitem{ap-04} G.E. Astrakharchik and L.P. Pitaevskii, Phys. Rev. A {\bf 70,}
013608 (2004).

\bibitem{kp-1970} B.B. Kadomtsev and V.I. Petviashvili, Sov. Phys. Doklady,
{\bf 15,} 539 (1970).

\bibitem{zakharov-1975} V.E. Zakharov, JETP Lett, {\bf 22,} 172 (1975).

\bibitem{aps-1997} J.C. Alexander, R.L. Pego, and R.L. Sachs,
Phys. Lett. A {\bf 226,} 187 (1997).

\bibitem{kt-1988} E.A. Kuznetsov and S.K. Turitsyn, Sov. Phys. JETP, {\bf 67,} 1583 (1988).

\bibitem{ll-10} E.M. Lifshitz and L.P. Pitaevskii, {\it Physical Kinetics,} (Pergamon, London, 1981).


\end{thebibliography}
\end{document}